\documentclass[preprint,english,showpacs,11pt,floatfix]{revtex4}

\usepackage{babel,amsmath,amssymb,dcolumn}
\usepackage[dvips]{graphics}

\begin{document}

\title{Slowly rotating proto strange stars in Quark mass
density- and temperature- dependent model}

\author{Jianyong Shen, Yun Zhang, Bin Wang}
\email{wangb@fudan.edu.cn} \affiliation{Department of Physics,
Fudan University, Shanghai 200433, People's Republic of China }

\author{Ru-Keng Su}
\email{rksu@fudan.ac.cn} \affiliation{China Center of Advanced
Science and Technology (World Laboratory), P.B.Box 8730, Beijing
100080, People's Republic of China
\\Department of Physics, Fudan University, Shanghai 200433,
People's Republic of China }

\begin{abstract}
Employing the Quark Mass Denisity- and temperature- dependent
model and the Hartle's method, We have studied the slowly rotating
strange star with uniform angular velocity. The mass-radius
relation, the moment of inertia and the frame dragging for
different frequencies are given. We found that we cannot use the
strange star to solve the challenges of Stella and Vietri for the
horizontal branch oscillations and the moment of inertia
$I_{45}/(M/M_ s)>2.3$. Furthermore, we extended the Hartle's
method to study the differential rotating strange star and found
that the differential rotation is an effective way to get massive
strange star.
\end{abstract}


\maketitle

\section{ Introduction}
The possible existence of strange quark stars which are made
entirely of u, d and s quarks is one of the most exciting aspects
of modern astrophysics\cite{s1}-\cite{s2}. It has been conjectured
that the compact objects such as X-ray pulsar Her X-1, X-ray
burstar 4U1820-30 are likely strange stars\cite{s3}-\cite{s5}. In
general, perhaps pulsars can be divided into two groups: one is
modeled as neutron stars, and the other as strange stars. It is
essential to distinguish these two candidates of pulsars from
observations and to predict their properties furthermore.

Usually, if strange stars do exist, the basic differences between
strange stars and neutron stars are their dynamical properties and
thermodynamical behaviors. For example, firstly, the mass-radius
(M-R) relation of strange stars is quite different from that of
neutron stars\cite{s5}, especially, this relation depends on the
rotation frequency of compact object. Secondly, for rotating
stars, the moment of inertia plays an important role to
investigate their dynamical and electromagnetic behavior. The
frequency of radio signals emitted from pulsars contain much
information on the moment of inertia of the sources, which helps
us not only to identify the type of pulsars -- neutron stars or
strange stars, but also to understand some phenomena and
mechanism. It is generally accepted that the sudden spin-ups,
glitches of pulsars, can be explained from their angular momentum
transfer between  their crusts and inner fluid. Obviously, this
transfer depends on their moments of inertia \cite{s6}. Thirdly, a
newly born neutron star or strange star may pass through various
stages of early evolution. These stages are determined by neutrino
time scale and this scale is proportional to the star radius and
the neutrino mean free path \cite{s7}. These two physical
quantities are dominated by temperature and the equation of state
(EOS) remarkably. In fact, due to the considerable difference
between the EOS of the strange quark matter and the neutron
matter, we can distinguish the strange star and the neutron star
from observation. The EOS at finite temperature plays the key role
to understand the behavior and the evolution of compact objects.

The early calculations of EOS for strange star are based on the
MIT bag model \cite{s8}. Later, many other models such as vector
interaction and density-dependent scalar potential model
\cite{s9}, the quark mass density-dependent (QMDD) model
\cite{s10}\cite{s11} and etc. were employed.

The QMDD model was first suggested by Fowler, Rata and
Weiner\cite{s12} and then used by many authors to study the
stability and the thermodynamical properties of strange quark
matter \cite{s13}-\cite{s16}. The basic hypothesis of the QMDD
model is that the masses of u, d quarks and strange quarks (and
the corresponding anti-quarks) are given by
\begin{equation}\label{1}
m_q  = \frac{B}{{3n_B }}\quad  (q = u,d,\bar u,\bar d),
\end{equation}
\begin{equation}\label{2}
m_{s,\bar s}  = m_{s0}  + \frac{B}{{3n_B }},
\end{equation}
where $n_B$ is the baryon number density, $m_{s0}$ is the current
mass of the strange quark and $B$ is the vacuum energy density.
Eqs.(\ref{1}) and (\ref{2}) can easily be understood from the
quark confinement mechanism, if we notice that the density $n_B$
approaches to zero and the masses of quarks tend to infinite when
the volume of the system goes infinite \cite{s17}. This
confinement mechanism is almost the same as that of the MIT bag
model. It has been shown by many authors that the thermodynamical
properties given by QMDD model are similar to that of MIT bag
model \cite{s14}, and the M-R relation, viscosity radial
oscillation of both rotating and non-rotating strange quark star
described by QMDD model are qualitatively similar to those
obtained with MIT bag model \cite{s10}.

Although the QMDD model can provide a dynamical description of
confinement and explain many aspects of strange quark matter, it
suffers from a basic drawback that it cannot reproduce a correct
lattice QCD deconfinement phase diagram because the quark masses
are divergent when $n_B \to 0$. To excite an infinite weight
particle, one must pay the price for infinite energy, i.e.
infinite temperature. It means that the QMDD model is a permanent
quark confinement model. It is unable to describe the
deconfinement process from neutron matter to quark matter inside
the pulsar. To overcome this difficulty, in a series of previous
papers \cite{s17}-\cite{s20} we suggested a quark mass density-
and temperature- dependent (QMDTD) model which is based on  a
non-permanent quark confinement Friedberg-Lee model and proved
that not only the above difficulty can be overcome but also many
physical properties, for example, the stability, the deconfinement
phase transition of strange quark matter\cite{s17}-\cite{s19}, the
binding energy of dibaryon system can be explained \cite{s20}.
Employing QMDTD model, Gupta and his co-workers studied the M-R
relation and the radial oscillations of proto strange stars and
found a lot of reasonable and interesting results \cite{s21}. But
their study was limited to non-rotating proto strange stars only.

This paper evolves from an attempt to extend the study of Gupta
et. al. to slowly rotating proto strange stars by using the QMDTD
model. We will employ the Hartle's method\cite{s22} to sketch the
main features about strange stars rotating with the frequency much
smaller than the Keplerian limit. We will discuss two cases: (1)
Uniform rotation where angular velocity $\Omega = const.$; (2)
$\Omega$ is a function of $r$, $\Omega = \Omega(r)$, which will be
called differential rotation below. We will investigate the
temperature dependence of M-R relation, of the moment of inertia
and of the effect of frame dragging for proto rotating strange
star.

The organization of this paper is as follows. In the following
section, we will give a brief review of the QMDTD model. By means
of the Hartle's formalism, we will discuss the slowly rotating
strange star with uniform angular velocity $\Omega$ in Sec.III,
and that with differential rotation in Sec.IV, respectively. The
last section contains a summary.

\section{The quark mass density- and temperature- dependent model}
We will give a brief review of the QMDTD model in this section.
The detail of this model can be found in
refs.\cite{s17}-\cite{s19}. Here we only write down the main steps
which are necessary for calculating the EOS for strange quark
matter.

According to the QMDTD model, the masses of u, d and s quarks (and
the corresponding anti-quarks) are given by
\begin{equation}\label{3}
m_q  = \frac{{B(T)}}{{3n_B }}\quad (q = u,d,\bar u,\bar d),
\end{equation}
\begin{equation}\label{4}
m_{s,\bar s}  = m_{s0}  + \frac{{B(T)}}{{3n_B }}.
\end{equation}
Comparing Eqs.(\ref{3}) and (\ref{4}) with Eqs.(\ref{1}) and
(\ref{2}), the constant $B$ in QMDD model is replaced by
\begin{equation}\label{5}
B(T) = B_0 [1 - (T/T_c )^2 ],\quad 0 \le T \le T_c
\end{equation}
\begin{equation}\label{6}
B(T) = 0,\quad \quad \quad \quad \quad \quad \;\;T > T_c
\end{equation}
where $B_0$ is the bag constant at zero temperature and $T_c = 170
MeV$ is the critical temperature of quark deconfinement phase
transition. The basic extension of QMDTD model is that $B(T)$
depends on temperature because according to the Friedberg-Lee
soliton bag model, the vacuum energy density of the bag equals the
different value between the local false vacuum minimum and the
absolute real vacuum minimum and this value depends on temperature
\cite{s23}.

\begin{figure}
\resizebox{0.7\linewidth}{!}{\includegraphics*{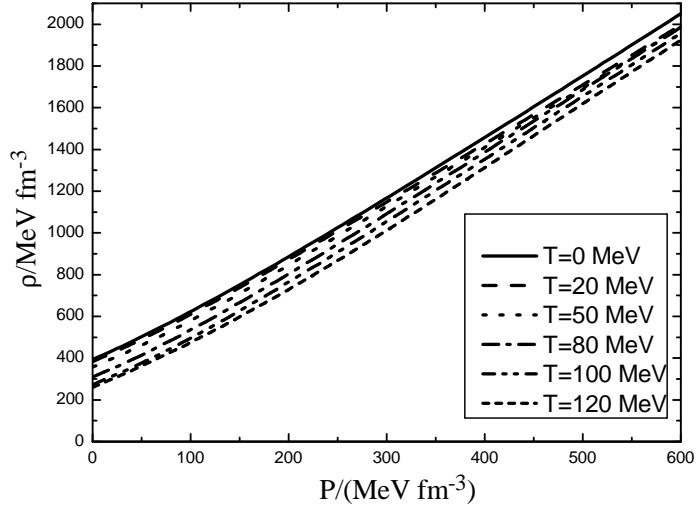}}
\caption{Diagram for the EOS of the strange matter in QMDTD
model.}
\label{f1}
\end{figure}

The thermodynamic potential reads
\begin{equation}\label{7}
Z =  - \sum\limits_i {T\int_0^\infty  {dk\frac{{dN_i (k)}}{{dk}}}
} \ln (1 + e^{ - \beta (\varepsilon _i (k) - \mu_i )} ),
\end{equation}
where $i$ stands for u, d, s (or $ \bar u,\bar d,\bar s $) quarks,
$\mu_i$ is the corresponding chemical potential (for anti-quark
$\mu_{\bar i}= -\mu_i$), $ \varepsilon _i  = \sqrt {m_i^2  + k^2
}$ is the single particle energy and $m_i$ is the mass for quarks
and anti-quarks. $ {{dN_i (k)} \mathord{\left/
 {\vphantom {{dN_i (k)} {dk}}} \right.
\kern-\nulldelimiterspace} {dk}} $ is the density of states for
various flavour quarks. The density of states of a spherical star
has been calculated in ref.\cite{s24}
\begin{equation}\label{8}
N_i (k) = A_i (kR)^3  + B_i (kR)^2  + C_i (kR),
\end{equation}
\begin{equation}\label{9}
A_i  = \frac{{2g_i }}{{9\pi }},
\end{equation}
\begin{equation}\label{10}
B_i  = \frac{{g_i }}{{2\pi }}\{ [1 + (\frac{{m_i }}{k})^2 ]\tan ^{
- 1} (\frac{k}{{m_i }}) - (\frac{{m_i }}{k}) - \frac{\pi }{2}\},
\end{equation}
\begin{eqnarray}\label{11}
C_i  &=& \frac{{g_i }}{{2\pi }}\{ \frac{1}{3} + (\frac{k}{{m_i }}
+ \frac{{m_i }}{k})\tan ^{ - 1} (\frac{k}{{m_i }}) - \frac{{\pi
k}}{{2m_i }} \nonumber\\
&&+ (\frac{{m_i }}{k})^{1.45} \frac{{g_i }}{{3.42(\frac{{m_i }}{k}
- 6.5)^2  + 100}}\},
\end{eqnarray}
where $g_i$ is the total degeneracy. For a slowly rotating star,
as a zero-th order approximation of the Hartle's perturbation
formalism, we use the density of states of a spherical cavity to
calculate.

After getting the thermodynamic potential $Z$, we can derive the
energy density $\rho$ and the pressure $p$. The results are
\cite{s15}\cite{s17}-\cite{s20}
\begin{equation}\label{12}
p =  - \frac{1}{V}\left. {\frac{{\partial (\Omega /n_B
)}}{{\partial (1/n_B )}}} \right|_{T,\mu _i }  =  - \frac{\Omega
}{V} + \frac{{n_B }}{V}\left. {\frac{{\partial \Omega }}{{\partial
n_B }}} \right|_{T,\mu _i },
\end{equation}
\begin{equation}\label{13}
\rho  = \frac{\Omega }{V} + \sum\limits_i {n_i \mu _i  -
\frac{T}{V}\left( {\frac{{\partial \Omega }}{{\partial T}}}
\right)} _{\mu _i ,n_B }.
\end{equation}
The extra terms in Eqs.(\ref{12}) and (\ref{13}) come from the
dependence of the quark mass on the baryon density.

The EOS for strange quark star can be calculated from above
formula and the result is shown in Fig.\ref{f1}, where the
parameters are fixed as $B_0=170 MeV fm^{-3}$, $m_{s0}=150MeV$ and
$T_c=170 MeV$.

\section{ Uniform rotating strange stars}
\subsection{Formalism}
In this section, we employ Hartle's formalism to study the slowly
rotating strange star. We review this method briefly below and the
details can be found in ref.\cite{s22}.

The metric of non-rotating configuration is
\begin{equation}\label{14}
 ds^2  =  - e^{\nu (r)} dt^2  +%
e^{\lambda (r)} dr^2  + r^2 (d\theta ^2  + \sin ^2 \theta d\varphi
^2 ).
\end{equation}
In hydrostatic equilibrium, $\nu(r)$ and $\lambda(r)$ are
determined by the TOV equations
\begin{equation}\label{15}
\frac{{d\nu }}{{dr}} =  - \frac{2}{{\rho  + P}}\frac{{dP}}{{dr}},
\end{equation}
\begin{equation}\label{16}
e^{\lambda (r)}  = (1 - \frac{{2m}}{r})^{ - 1},
\end{equation}
\begin{equation}\label{17}
\frac{{dm}}{{dr}} = 4\pi r^2 \rho,
\end{equation}
\begin{equation}\label{18}
\frac{{dP}}{{dr}} = \frac{{(P + \rho )(m + 4\pi r^3 P)}}{{r^2 (1 -
\frac{{2m}}{r})}},
\end{equation}
where $P$ and $\rho$ are the pressure and energy density of
strange matter respectively and they relate each other by EOS. The
stellar mass $M=m(R)$, where $R$ is the radius of the star and is
obtained by solving equation $P(R)=0$. For a slowly rotating star,
the configuration is no longer static and spherically symmetric,
but axially symmetric instead. The metric becomes
\begin{eqnarray}\label{19}
ds^2  &=&  - e^{\nu (r)} (1 + 2h)dt^2 + e^{\lambda (r)} [1 +
\frac{{2\tilde m}}{{r - 2m}}]dr^2 \nonumber \\
&&+ r^2 (1 + 2k)[d\theta ^2  + \sin ^2 \theta (d\varphi  - \omega
dt)^2 ] + O(\omega ^3 ),
\end{eqnarray}
where $\omega(r)$ is the angular velocity of a locally
non-rotating inertial frame. Expanding the metric in spherical
harmonics, we find
\begin{equation} \label{20}
\left\{ {\begin{array}{*{20}c}
   {h = h(r,\theta ) = h_0 (r) + h_1 (r)P_1 (\theta ) + h_2 P(\theta ) + ...}  \\
   {\tilde m = \tilde m(r,\theta ) = \tilde m_0 (r) + \tilde m_1 (r)P_1 (\theta ) + \tilde m_2 P(\theta ) + ...}  \\
   {k = k(r,\theta ) = k_0 (r) + k_1 (r)P_1 (\theta ) + k_2 P(\theta ) + ...}  \\
\end{array}} \right.
\end{equation}
It can be proved from symmetric consideration that $h_1(r)$,
$\tilde m_1(r)$ and $k_1(r)$ vanish. Therefore, under the slow
rotation approximation the metric can be written as
\begin{eqnarray}\label{21}
ds^2  &=&  - e^\nu  [1 + 2(h_0  + h_2 P_2 )]dt^2  + e^{\lambda }
[1 + \frac{{2(\tilde m_0  + \tilde m_2 P_2 )}}{{r - 2m}}] dr^2
\nonumber \\
&& + r^2 (1 + 2k_2 P_2 )[d\theta ^2  + \sin ^2 \theta(dt - \omega
d\varphi )^2 ] + O(\omega ^3
 ),
\end{eqnarray}
where $e^\nu$ and $e^\lambda$ can be calculated by TOV equations
(\ref{15})-(\ref{18}) and $P_2=(3 cos \theta -1)/2$.

The rotation not only affects the metric, but also can change the
distribution of pressure in the interior of the star. In a
co-moving reference frame with the fluid, the pressure correction
is
\begin{equation}\label{22}
P + (P + \rho )p^* = P + \Delta P,
\end{equation}
and the energy density correction
\begin{equation}\label{23}
\rho  + (P + \rho )\frac{{d\rho }}{{dP}}p^* = \rho  + \Delta \rho.
\end{equation}
Here $p^*$ indicates the dimensionless corrections on the
pressure. It can also be expanded as
\begin{equation}\label{24}
p^*(r, \theta)=p_0^*(r)  + p_2^*(r) P_2 + ...
\end{equation}
The stress-energy tensor in a rotating star is
\begin{equation}\label{25}
T^{(0)\nu } _\mu   + \Delta T^\nu  _\mu   =  (\rho  + \Delta \rho
+ P + \Delta P)u_\mu  u^\nu   + (P + \Delta P)\delta _\mu ^\nu,
\end{equation}
where
\begin{equation}\label{26}
\left\{ \begin{array}{l}
 u^t  = ( - g_{tt}  - 2\Omega g_{t\varphi }  - g_{\varphi \varphi } \Omega ^2 )^{ - 1/2}  \\
 u^\varphi   = \Omega u^t ,u^r  = u^\theta   = 0 \\
 \end{array} \right.
.\end{equation} $\Omega$ is the angular velocity of fluid observed
far from the star. Noting that the perturbed Einstein equation
$G^{(0)\nu}_\mu + \Delta G^\nu _\mu = 8\pi T^{(0)\nu }_\mu + 8\pi
\Delta T^\nu _\mu$ and the zeroth order equations $G^{(0)\nu
}_\mu=8\pi T^{(0)\nu }_\mu$, we get
\begin{equation}\label{27}
\Delta G^\nu _\mu = 8\pi \Delta T^\nu _\mu.
\end{equation}

For the component of $t = \nu$ and $\mu = \phi$ of Eq.(\ref{27}) ,
we have
\begin{eqnarray}\label{28}
 \frac{1}{{r^4 }}\frac{\partial }{{\partial r}}\{ r^4 j(r)\frac{{\partial \varpi }}{{\partial r}}\}  + \frac{4}{r}\frac{{dj}}{{dr}}\varpi  + \frac{{e^{ - (\nu  - \lambda )/2} }}{{r^2 \sin ^3 \theta }}\frac{\partial }{{\partial \theta }}\{ \sin ^3 \theta \frac{{\partial \varpi }}{{\partial \theta }}\} \nonumber \\
  = \frac{1}{{r^4 }}\frac{\partial }{{\partial r}}\{ r^4 j(r)\frac{{\partial \Omega }}{{\partial r}}\}  + \frac{{e^{ - (\nu  - \lambda )/2} }}{{r^2 \sin ^3 \theta }}\frac{\partial }{{\partial \theta }}\{ \sin ^3 \theta \frac{{\partial \Omega }}{{\partial \theta }}\},
\end{eqnarray}
where
\begin{equation}\label{29}
j(r) = e^{ - (\nu  + \lambda )/2}
\end{equation}
and we have defined $\varpi=\Omega-\omega$. For the uniform
rotation, the angular velocity of fluid is a constant
$\Omega=const.$. Therefore, the right hand side of Eq.(\ref{28})
vanishes. We expand $\varpi(r, \theta)$ \cite{s25}:
\begin{equation}\label{30}
\varpi (r,\theta ) = \sum\limits_{l = 1}^\infty  {\varpi _l (r)( -
\frac{1}{{\sin \theta }}\frac{{dP_l }}{{d\theta }})}.
\end{equation}
Then the radial function of ${\varpi _l (r)}$ satisfies
\begin{equation}\label{31}
\frac{1}{{r^4 }}\frac{d}{{dr}}\{ r^4 j(r)\frac{{d\varpi _l
}}{{dr}}\}  + \{ \frac{4}{r}\frac{{dj}}{{dr}} - e^{ - (\nu  -
\lambda )/2} \frac{{l(l + 1) - 2}}{{r^2 }}\} \varpi _l  = 0.
\end{equation}
At large $r$, ${\varpi _l (r)}$ has the form
\begin{equation}\label{32}
\varpi _l (r) \to const.r^{ - l - 2}  + const.r^{l - 1}.
\end{equation}
At infinity where the spacetime is flat, $\omega$ must decrease
faster than $1/r^3$. Because of $\varpi  = \Omega  - \omega$, only
the coefficient of $l=1$ in the $\varpi$ expansion does not vanish
and Eq.(\ref{31}) reduces to
\begin{equation} \label{33}
\frac{1}{{r^4 }}\frac{d}{{dr}}\{ r^4 j(r)\frac{{d\varpi }}{{dr}}\}
+ \frac{4}{r}\frac{{dj}}{{dr}}\varpi  = 0,
\end{equation}
where $\varpi=\varpi_1(r)$. Outside the star $j(r)=1$, the
solution reads
\begin{equation}\label{34}
\varpi (r) = \Omega  - \frac{{2J}}{{r^3 }},
\end{equation}
where the constant $J$ is identified as the total angular momentum
of star and satisfies
\begin{equation}\label{35}
J = \int {T_\varphi ^t \sqrt { - {}^3g} } dV,
\end{equation}
where ${}^3g$ is the determinant of metric of space components.
The moment of inertia is simply
\begin{equation}\label{36}
I = {\raise0.7ex\hbox{$J$} \!\mathord{\left/
 {\vphantom {J \Omega }}\right.\kern-\nulldelimiterspace}
\!\lower0.7ex\hbox{$\Omega $}}.
\end{equation}

\begin{figure}
\resizebox{0.7\linewidth}{!}{\includegraphics*{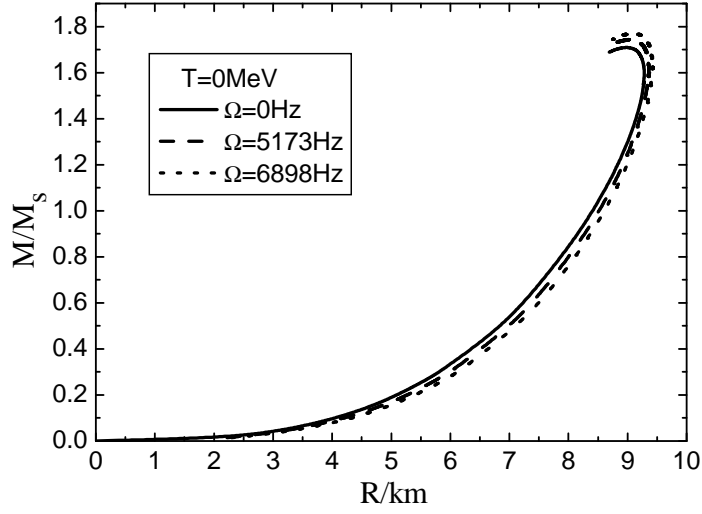}}
\caption{The radius $R$ in km vs. mass $M/M_s$ at the temperature
$T=0MeV$ for different angular velocity $\Omega$.}
\label{f2a}
\end{figure}

\begin{figure}
\resizebox{0.7\linewidth}{!}{\includegraphics*{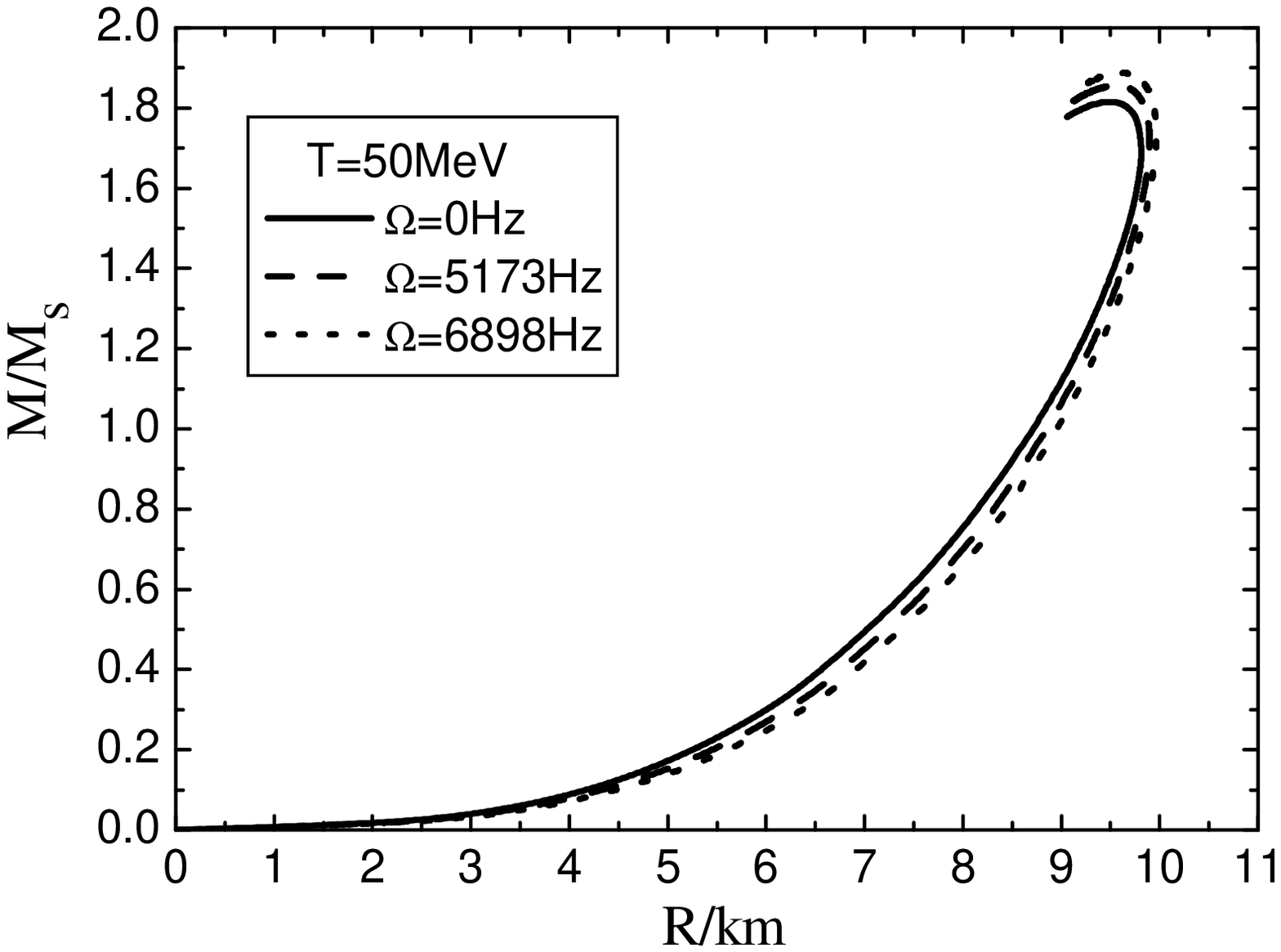}}
\caption{The radius $R$ in km vs. mass $M/M_s$ at the temperature
 $T=50MeV$ for different angular velocity $\Omega$.}
\label{f2b}
\end{figure}

The components $\nu=\mu=t$ and $\nu=\mu=r$ of Eq.(\ref{27}) and
the terms $l=0$ in the expansion (\ref{20}) lead to
\begin{equation}\label{37}
\frac{{d\tilde m_0 }}{{dr}} = 4\pi r^2 p_0^* (\rho  +
P)\frac{{d\rho }}{{dP}} + \frac{1}{{12}}j^2 r^4 \left(
{\frac{{d\omega }}{{dr}}} \right)^2  - \frac{1}{3}r^3 \frac{{dj^2
}}{{dr}}\varpi ^2,
\end{equation}
\begin{equation}\label{38}
\frac{{dh_0 }}{{dr}} - \frac{{\tilde m_0 r^2 }}{{(r - 2m)^2
}}(8\pi P + \frac{1}{{r^2 }}) = \frac{{4\pi (\rho  + P)r^2 }}{{r -
2m}}p_0^* - \frac{1}{{12}}\frac{{r^2 }}{{r - 2m}}j^2 \left(
{\frac{{d\omega }}{{dr}}} \right)^2.
\end{equation}
Another constraint between the perturbations of the metric and
those of energy and pressure comes from the hydrostatic
equilibrium, i.e. the conservation of stress-energy tensor of
perfect fluid $T^{\mu \nu }_{;\nu}=0$. In the non-rotating case,
it can be expressed as
\begin{equation} \label{39}
\frac{{p_{,i} }}{{p + \rho }} - \frac{{\nu _{,i} }}{{2\nu }} = 0
\end{equation}
and in the rotating case, the corresponding perturbation equation
reads
\begin{equation}\label{40}
p^* _{,i}  + h_{,i}  - \frac{1}{2}(e^{ - \nu } \varpi ^2 r^2 \sin
^2 \theta )_{,i}  = 0.
\end{equation}
For the terms $l=0$, we find
\begin{equation}\label{41}
\frac{{dp_0^* }}{{dr}} + \frac{{dh_0 }}{{dr}} -
\frac{1}{3}\frac{d}{{dr}}(e^{ - \nu } \varpi ^2 r) = 0.
\end{equation}
Combining Eqs.(\ref{41}) and (\ref{38}), we find
\begin{eqnarray}\label{42}
 - \frac{{dp_0^* }}{{dr}} + \frac{1}{{12}}\frac{{r^2 }}{{r - 2m}}j^2 \left( {\frac{{d\omega }}{{dr}}} \right)^2  + \frac{1}{3}\frac{d}{{dr}}(\frac{{r^3 j^2 \varpi ^2 }}{{r -
 2m}}) \nonumber \\
 = \frac{{4\pi (\rho  + P)r^2 }}{{r - 2m}}p_0^*  + \frac{{\tilde m_0 r^2 }}{{(r - 2m)^2 }}(8\pi P + \frac{1}{{r^2
 }}).
\end{eqnarray}
On the other hand, one can prove that the increasing of stellar
mass $\delta M$ and mean radius $\delta R$ are given by \cite{s22}
\begin{equation}\label{43}
\delta M = \tilde m_0 (R) + \frac{{J^2 }}{{R^3 }},
\end{equation}
\begin{equation}\label{44}
\delta R = p_0^* (P + \rho )/\left( {\frac{{dP}}{{dr}}} \right).
\end{equation}
When EOS, $\Omega$ and the boundary value of $P(0)$ are given, we
can integrate the TOV equations until reaching the surface of the
star $P(R)=0$. Then we solve the Eq.(\ref{33}), Eq.(\ref{37}) and
Eq.(\ref{42}) to get the perturbations and finally calculate the
moment of inertia by Eq.(\ref{36}), the correction of the mass and
radius in rotational case by Eq.(\ref{43}) and Eq.(\ref{44}). We
do this numerical calculation with the rotational frequencies
smaller than the Keplerian frequency
\begin{equation}\label{44.5}
\Omega ^2 < \Omega _k^2 \sim M/R^3,
\end{equation}
which is required by the Hartle's method\cite{s22}.

\subsection{Results}
\subsubsection{M-R curves}
\begin{figure}
\resizebox{0.7\linewidth}{!}{\includegraphics*{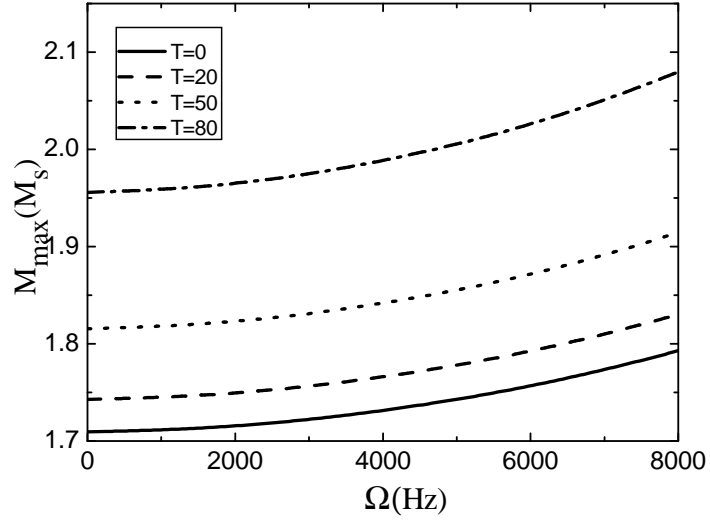}}
\caption{The increase of maximum mass $\delta M_{max}/M_s$ vs. the
angular velocity $\Omega$ at different temperatures.}
\label{f3}
\end{figure}

The Mass-Radius relation is one of the main results which
distinguishes the strange stars and the neutron stars. In our
calculation, the M-R relations not only depend on temperature $T$,
but also on angular velocity $\Omega$. Fix the temperature $T=0$
and $50 MeV$, the M-R curves for different $\Omega$ are shown in
Fig.\ref{f2a} and \ref{f2b}, respectively. By using Fig.\ref{f2a},
Fig.\ref{f2b} and Eq.(\ref{44.5}), we find Keplerian frequency
$\Omega_k \approx 1.5 \times 10^4 Hz$, which is much larger than
the frequencies used in Fig.\ref{f2a} and Fig.\ref{f2b}. It is
seen that the maximum mass and the corresponding radius increase
with $\Omega$. To show the temperature effect more transparently,
we plot the $M_{max}$ vs. $\Omega$ curves for different
temperatures in Fig.\ref{f3}, and see that for a fixed $\Omega$,
$M_{max}$ increases with temperature. The effect of rotation
becomes more and more important when temperature increases. From
Fig.\ref{f3}, we can express the increase of the maximum mass by a
simple formula
\begin{equation}\label{45}
{\raise0.7ex\hbox{${\delta M_{\max } }$} \!\mathord{\left/
 {\vphantom {{\delta M_{\max } } {M_s }}}\right.\kern-\nulldelimiterspace}
\!\lower0.7ex\hbox{${M_s }$}} \propto \Omega ^2,
\end{equation}
where $M_s$ is the mass of the sun. This is very reasonable
because all perturbations in Hartle's method are assumed to be
proportional to the square of the angular velocity of the fluid.

\begin{figure}
\resizebox{0.7\linewidth}{!}{\includegraphics*{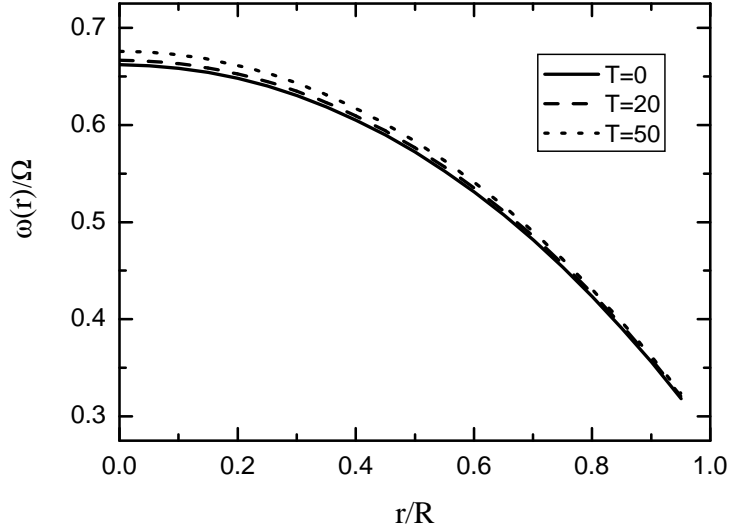}}
\caption{The ratio of angular velocity of inertial frame and fluid
in the interior of the star $\omega(r)/\Omega$ vs. $r/R$, where
$R$ is the radius of the star, at the temperatures $T=20,50 MeV$.}
\label{f4}
\end{figure}

\subsubsection{The effect of frame dragging}
The frame dragging is a general relativistic effect for a rotating
object. To show the effect of frame dragging in a slowly rotating
star, we plot the $\omega(r)/\Omega$ vs. $r/R$ curves for
different temperatures in Fig.\ref{f4}. We see from Fig.\ref{f4}
that this effect decreases when one approaches to the surface of
star, and the dragging increases when temperature increases.

\begin{figure}
\resizebox{0.7\linewidth}{!}{\includegraphics*{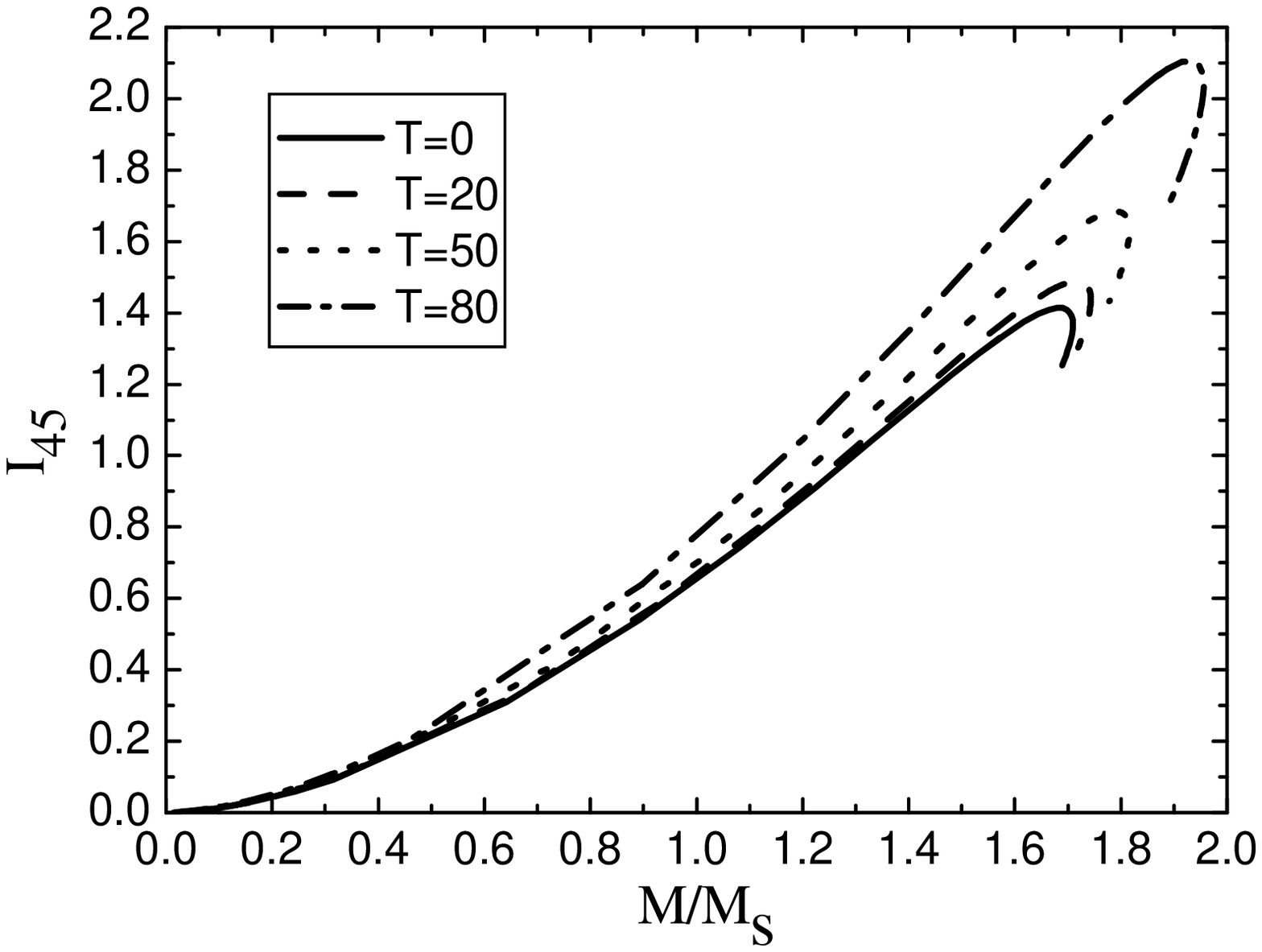}}
\caption{The moment of inertia $I_{45}$ vs. mass $M/M_s$ at the
different temperatures $T=0,20,50,80MeV$ for the star rotating at
the frequency $\Omega = 200Hz$.}
\label{f6}
\end{figure}

\subsubsection{The moment of inertia}
Besides the M-R relation, much information for the configuration
of a strange star is provided by its moment of inertia, which is
one of the essential facts to understand many observational
phenomenons. Fig.\ref{f6} shows the relation of the moment of
inertia-Mass. We see that the moment of inertia will increase with
temperature.

\begin{figure}
\resizebox{0.7\linewidth}{!}{\includegraphics*{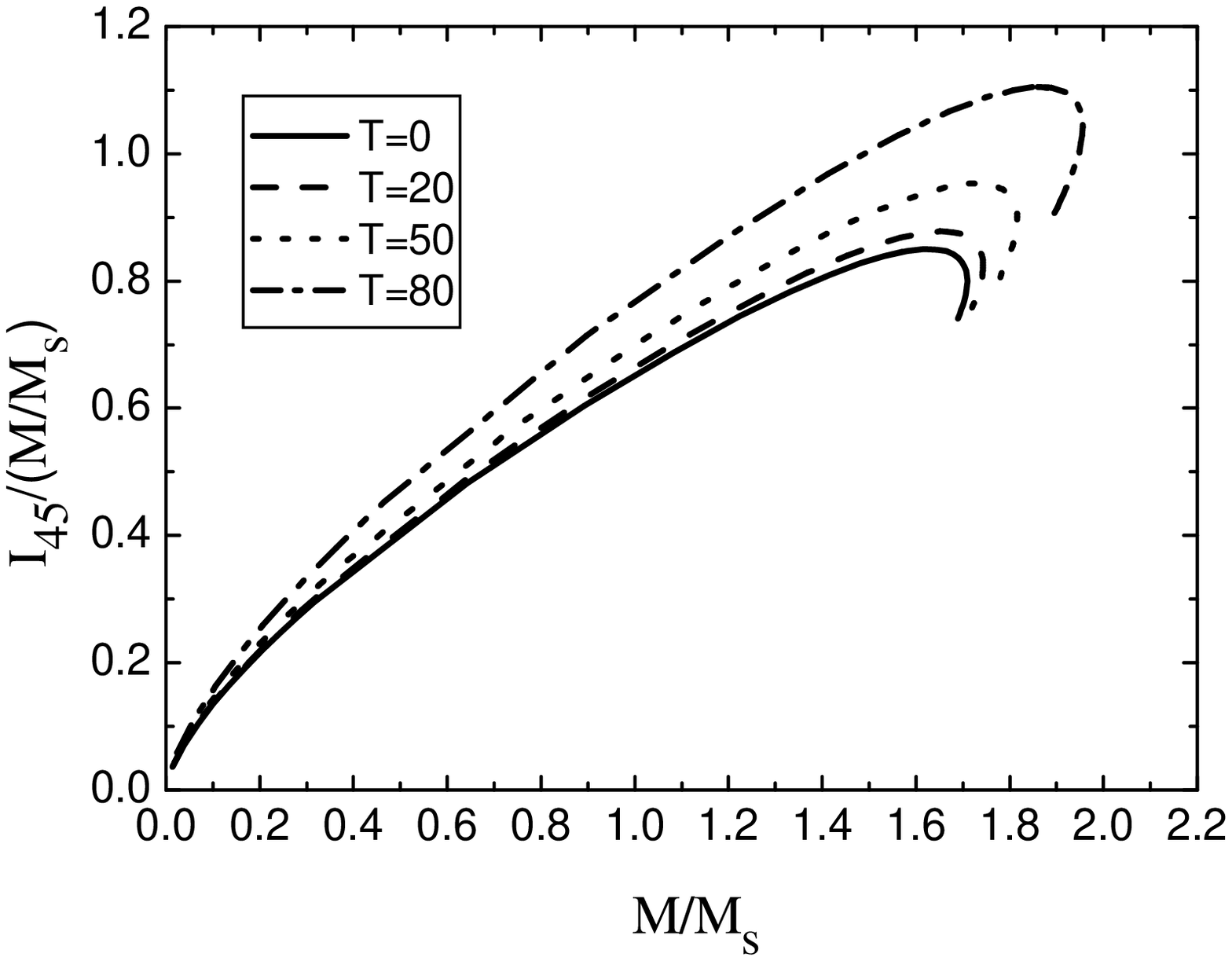}}
\caption{The ratio of the moment of inertia and the stellar mass
$I_{45}/(M/M_s)$ vs. mass $M/M_s$ at the different temperatures
$T=0,20,50,80MeV$.}
\label{f7}
\end{figure}

\subsubsection{Horizontal-branch oscillation}
Recently Stella and Vietri\cite{s30} made a link between
quasi-periodic oscillations(QPO) and general relativistic
Lense-Thirring precession caused by frame dragging. The low
frequency of QPO, called horizontal-branch oscillations(HBO), is
the frequency of the L-T precession of an inclined circular orbit
with Keplerian orbital frequency. Psaltis et al.\cite{s31} found
that the ratios of the moments of inertia and masses $ {{I_{45} }
\mathord{\left/
 {\vphantom {{I_{45} } {(\frac{M}{{M_s }})}}} \right.
 \kern-\nulldelimiterspace} {(\frac{M}{{M_s }})}}
$ of five bright sources (GX 17+2, Cyg X-2, GX 5-1, Sco X-1, GX
340+0) satisfy $I_{45}/(M/M_ s)>2.3$. Kalogera and
Psaltis\cite{s32} calculate this ratio for various models of
neutron star and argue that this requirement cannot be satisfied
for neutron stars.

Here we suppose these five sources are strange stars and discuss
whether the condition about $I_{45}/(M/M_ s)$ is satisfied.
Fig.\ref{f7} tells us that to satisfy the condition $I_{45}/(M/M_
s)>2.3$ one require too high temperature which is over $120MeV$
from the Figure. This makes the validity of the suggestion by
Stella and Vietri to be challenged also for the strange stars.

\section{Differential Rotation Strange Stars }
We will study the configuration of differential rotating star,
i.e. the angular velocity of the fluid $\Omega=\Omega(r,\theta)$,
in this section. Obviously, differential rotation will affect the
evolution of a hot and new-born star. Noting that Hartle's method
can only be used to study the uniform rotational star, an
extension is needed when star is newly born and differentially
rotating.

\subsection{Formulism}
It can easily be seen that the Eq.(\ref{28}) is still valid in the
case of differential rotation, but the right hand side does not
varnish now. Expanding $\Omega(r,\theta)$ (and then
$\varpi(r,\theta)$) by the Legendre functions
\begin{equation}\label{46}
\Omega (r,\theta ) = \sum\limits_{l = 1}^\infty  {\Omega _l (r)( -
\frac{1}{{\sin \theta }}\frac{{dP_l }}{{d\theta }})}.
\end{equation}
Eq.(\ref{28}) becomes
\begin{eqnarray}\label{47}
 \frac{1}{{r^4 }}\frac{d}{{dr}}\{ r^4 j(r)\frac{{d\varpi _l }}{{dr}}\}  + \{ \frac{4}{r}\frac{{dj}}{{dr}} - e^{ - (\nu  - \lambda )/2} \frac{{l(l + 1) - 2}}{{r^2 }}\} \varpi _l  \nonumber \\
  = \frac{1}{{r^4 }}\frac{d}{{dr}}\{ r^4 j(r)\frac{{d\Omega _l }}{{dr}}\}  - e^{ - (\nu  - \lambda )/2} \frac{{l(l + 1) - 2}}{{r^2 }}\Omega _l .
\end{eqnarray}
Remember that outside the star, there is no fluid,
$\Omega_l(r)=0$, therefore, $\varpi_l(r)$ in Eq.(\ref{47}) still
has the form
\[
\varpi _l (r) \to const.r^{ - l - 2}  + const.r^{l - 1}.
\]
A similar discussion can be made as that of the uniform rotation
case: $\omega$ should decrease faster than $1/r^3$ if the
spacetime is asymptotically flat. As a result, only the term with
$l=1$ remains, while $\varpi=\varpi_1(r)$ and $\Omega=\Omega_1(r)$
are the functions of radial coordinate only. Eq.(\ref{47}) is
simplified to be
\begin{equation}\label{48}
\frac{1}{{r^4 }}\frac{d}{{dr}}\{ r^4 j(r)\frac{{d\varpi }}{{dr}}\}
+ \frac{4}{r}\frac{{dj}}{{dr}}\varpi  = \frac{1}{{r^4
}}\frac{d}{{dr}}\{ r^4 j(r)\frac{{d\Omega }}{{dr}}\},
\end{equation}
with $\varpi'(0)=\Omega'(0)$. The solution is
\begin{equation}\label{49}
\varpi (r) = const. - \frac{{2J}}{{r^3 }}.
\end{equation}
Here the constant $J$ can still be interpreted as the total
angular momentum of the star. We obtain
\begin{equation}\label{185}
\varpi (R) = \Omega (R) - \frac{{2J}}{{R^3 }}
\end{equation}
at the surface of star $r=R$. In differential rotation, the
angular velocity of fluid depends on the radial coordinate. This
means that in the internal regime of the star, a very thin layer
$r \to r+dr$ of the fluid rotates with the same angular velocity
$\Omega(r)$ and generates angular momentum of this layer $dJ$,
which satisfies,
\[
dJ=\Omega(r)dI.
\]
$dI$ is the moment of inertia of this thin layer. Hence the total
moment of inertia of the star is
\begin{equation}\label{51}
I = \int {dI = \int {dJ/\Omega  = \int {({{T_\varphi ^t }
\mathord{\left/
 {\vphantom {{T_\varphi ^t } \Omega }} \right.
 \kern-\nulldelimiterspace} \Omega })\sqrt { - {}^3g} dV} } },
\end{equation}
where the Eq.(\ref{35}) is used at the last equality. Since the
differential rotation changes the derivatives of stress-energy
tensor in hydrostatic equilibrium, we find
\begin{equation} \label{52}
\frac{{p_{,i} }}{{p + \rho }} - \frac{{u_{,i}^t }}{{u^t }} +
F(\Omega )\Omega _{,i}  = 0,
\end{equation}
where
\begin{equation} \label{53}
F(\Omega ) =  - \frac{{g_{t\phi }  + \Omega g_{\phi \phi }
}}{{g_{tt}  + 2\Omega g_{t\phi }  + \Omega ^2 g_{\phi \phi } }}.
\end{equation}
In the slowly rotating approximation, the effect of differential
rotation is considered as perturbation. Therefore we have
\begin{equation} \label{54}
p_{,i}^*  + h_{,i}  - \frac{1}{2}(\varpi ^2 r^2 \sin ^2 \theta e^{
- \nu } )_{,i}  + (e^{ - \nu } \varpi r^2 \sin ^2 \theta )\Omega
_{,i}  = 0.
\end{equation}
For the term $l=0$ and $i=r$, Eq.(\ref{54}) gives
\begin{equation} \label{55}
p_{0,r}^*  + h_{0,r}  - \frac{1}{3}(\varpi ^2 r^2 e^{ - \nu }
)_{,r}  + \frac{2}{3}(e^{ - \nu } \varpi r^2 )\Omega _{,r}  = 0.
\end{equation}
Combining with Eq.(\ref{38}), we find
\begin{eqnarray}\label{56}
 - \frac{{dp_0^* }}{{dr}} &=& \frac{{4\pi (\rho  + p)}}{{r - 2m}}r^2 p_0^*  + \frac{{\tilde m_0 (8\pi pr^2  + 1)}}{{(r - 2m)}} - \frac{{r^4 j^2 }}{{12(r - 2m)}}(\frac{{d\omega
 }}{{dr}})^2 \nonumber \\
 && - \frac{1}{3}\frac{d}{{dr}}(\frac{{r^3 j^2 \varpi ^2 }}{{r - 2m}}) + \frac{2}{3}\frac{{r^3 j^2 \varpi }}{{r - 2m}}\frac{{d\Omega }}{{dr}},
\end{eqnarray}
with boundary conditions $\tilde m_0=p^*_0=0$. Eq.(\ref{37}) is
valid in the differential rotation, because the derivatives of
that $\Omega(r)$ do not appear in both of the right and left hand
sides of Eq.(\ref{27}) as in the uniform rotation. We can solve
Eq.(\ref{37})(\ref{56})numerically.

\begin{figure}
\resizebox{0.7\linewidth}{!}{\includegraphics*{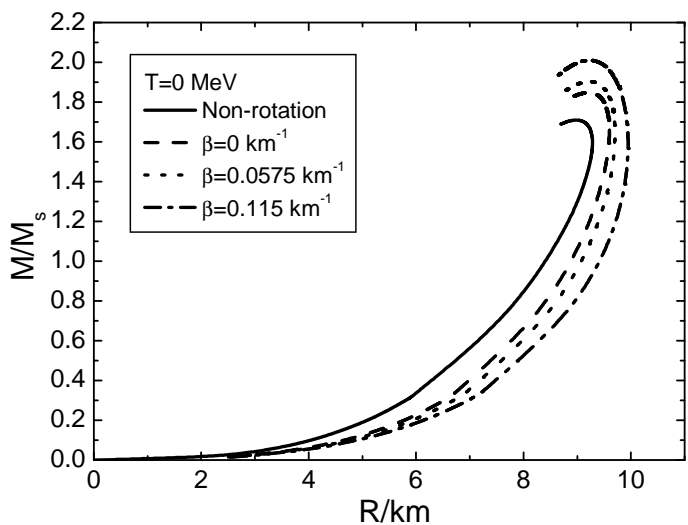}}
\caption{The radius $R$ vs. mass $M/M_s$ at the temperature
$T=0MeV$ for the different values of $\beta$.}
\label{f8a}
\end{figure}

\begin{figure}
\resizebox{0.7\linewidth}{!}{\includegraphics*{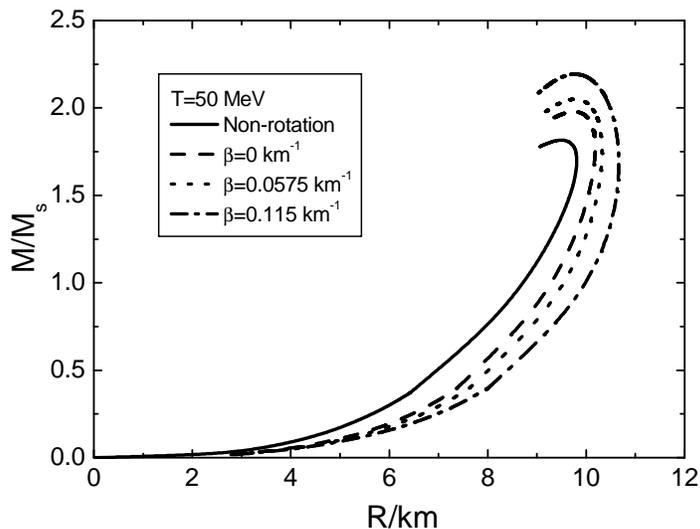}}
\caption{The radius $R$ vs. mass $M/M_s$ at the temperature
$T=50MeV$ for the different values of $\beta$. }
\label{f8b}
\end{figure}

\subsection{Results}
\begin{figure}
\resizebox{0.7\linewidth}{!}{\includegraphics*{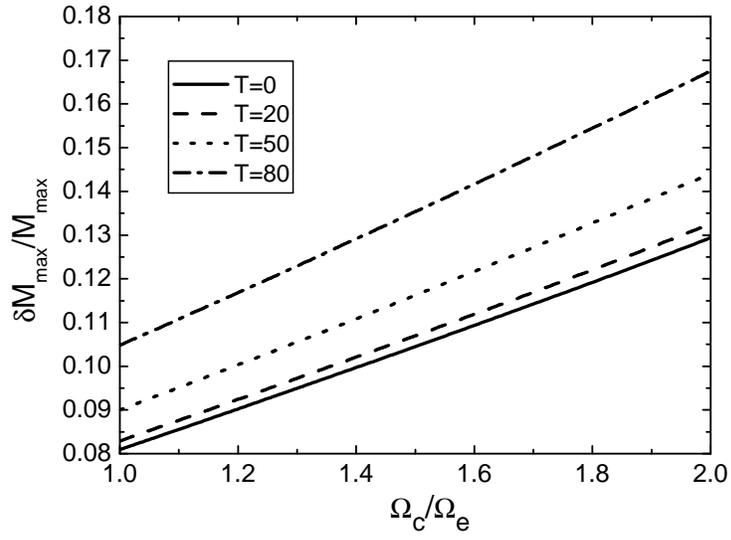}}
\caption{The ratio of the maximum mass increase $\delta
M_{max}/M_{max}$ vs. $\Omega_c/\Omega_e$, the ratio of the angular
velocity at the core and at the surface at the different
temperatures $T=0,20,50,80MeV$. }
\label{f9}
\end{figure}

We employ the distribution of the angular velocity of the fluid
\begin{equation}\label{57}
\Omega (r) = \Omega _c e^{ - \beta r},
\end{equation}
where $\Omega_c$ is the angular velocity of the core and $\beta$
is a constant, to investigate the influences of differential
rotation. Fixing the surface angular velocity $\Omega_e=10346 Hz$,
which is still smaller than the Mass-shedding limit at the surface
of a uniformly rotating star and satisfies the slowly rotating
approximation, we show the M-R curves for $\Omega(r)$ at
temperature $T=0$ and $50MeV$ in Fig.\ref{f8a} and Fig.\ref{f8b}
respectively. We see that the effect of $\beta$ on $M_{max}$ is
remarkable, and the result reduces to the uniform rotation when
$\beta=0$. A differentially rotating strange star can sustain more
mass than that of the uniform rotating configuration. The curves
of the maximum mass increase $\delta M_{max}/M_{max}$ vs.
$\Omega_c/\Omega_e$, the ratio of the angular velocity at the core
and at the surface, are shown in Fig.\ref{f9}. We see that $\delta
M_{max}/M_{max}$ changes considerably with temperature and
$\Omega_c/\Omega_e$. A hot and differentially rotating strange
star will suffer more maximum mass increase.

\begin{figure}
\resizebox{0.7\linewidth}{!}{\includegraphics*{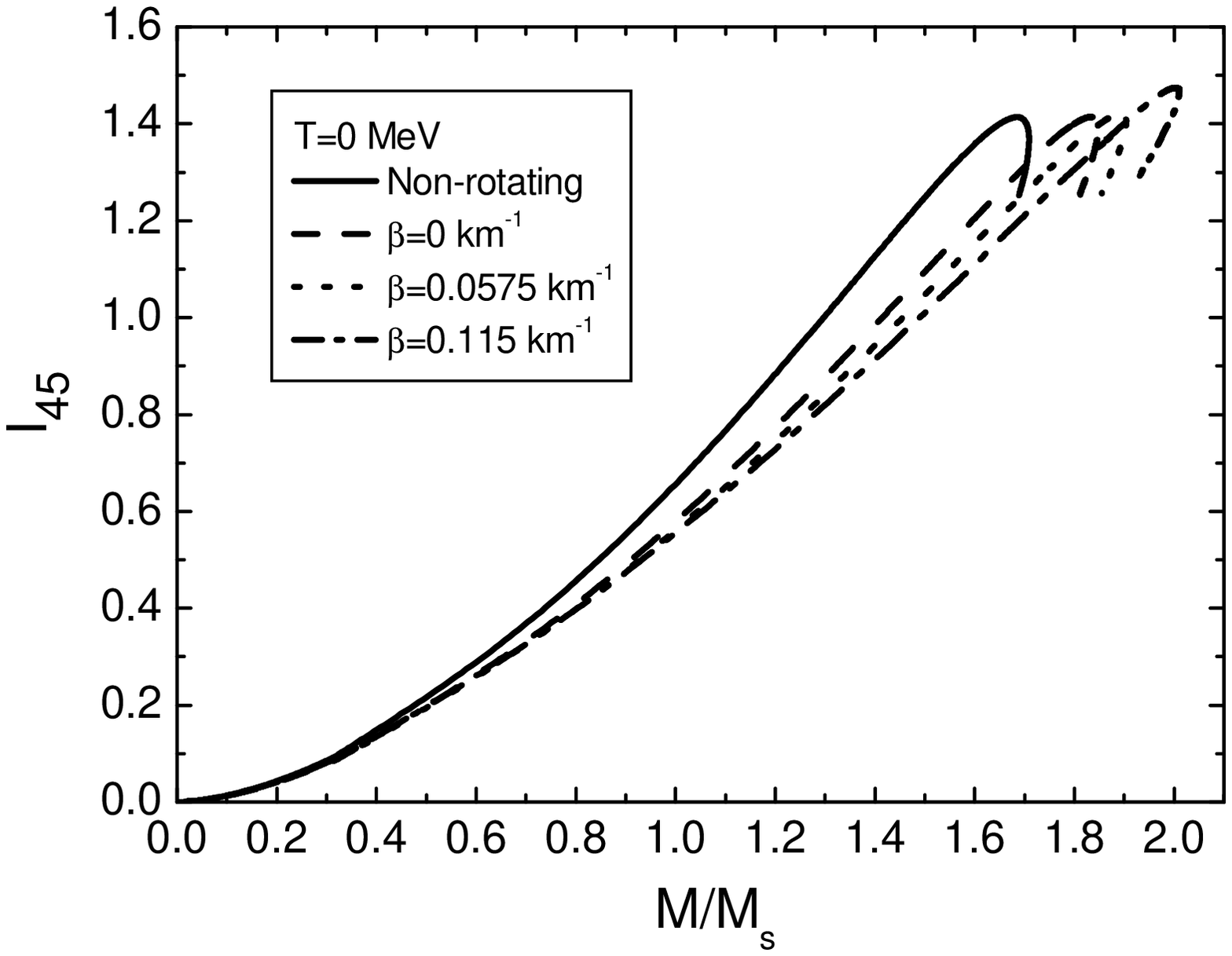}}
\caption{The moment of inertia $I_{45}$ vs. mass $M/M_s$ at the
temperature $T=0MeV$ for the different values of $\beta$. }
\label{f10a}
\end{figure}

\begin{figure}
\resizebox{0.7\linewidth}{!}{\includegraphics*{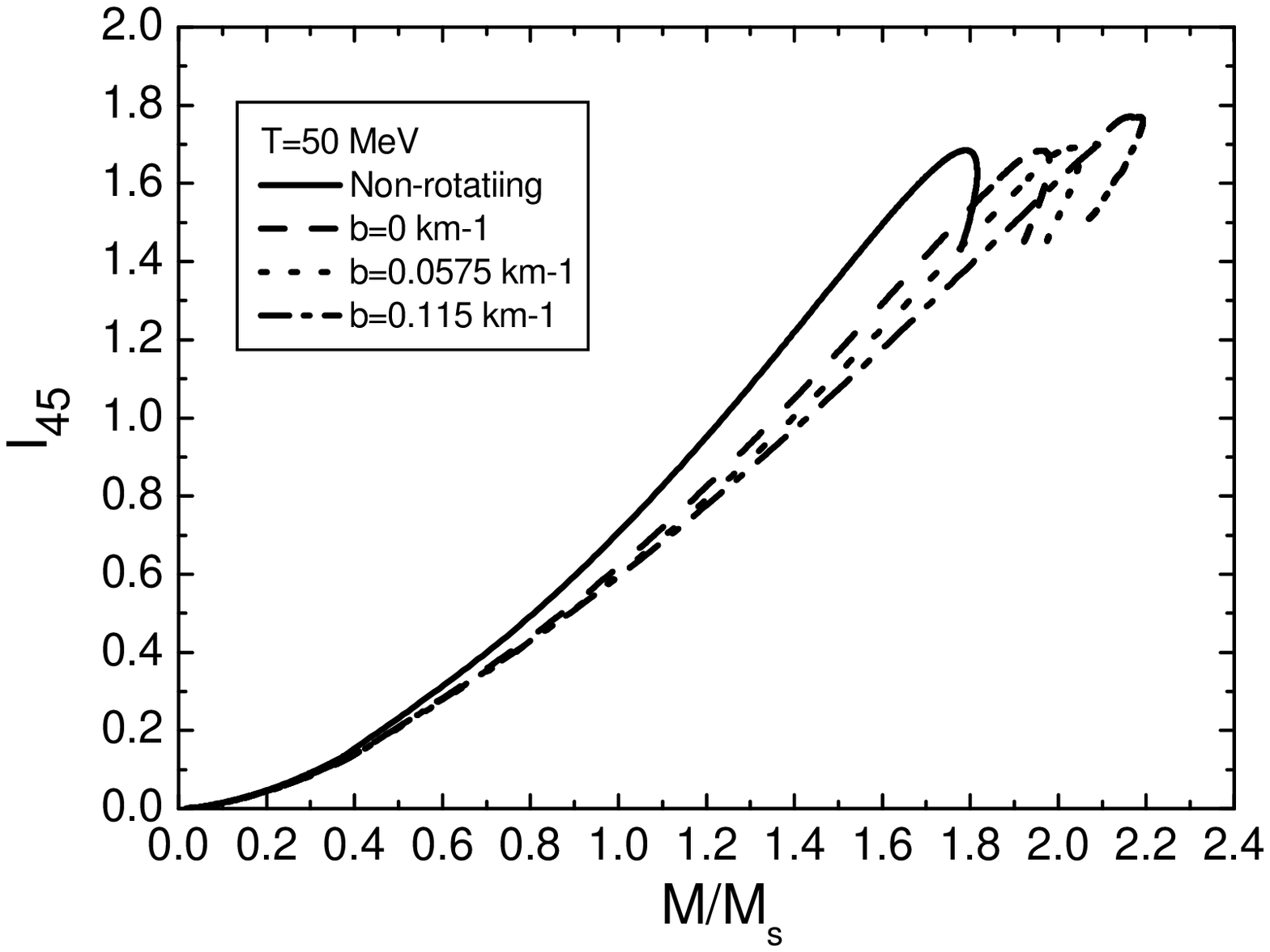}}
\caption{he moment of inertia $I_{45}$ vs. mass $M/M_s$ at the
temperature $T=50MeV$ for the different values of $\beta$. }
\label{f10b}
\end{figure}

Similarly, we can address the effect of $\Omega(r)$ on the moment
of inertia at temperature $T=0$ and $50MeV$. We show this effect
in Fig.\ref{f10a} and Fig.\ref{f10b} respectively. Comparing to
Fig.\ref{f8a} and Fig.\ref{f8b}, we find that corresponding to the
mass, the effect of $\beta$ on the moment of inertia is small.
This is reasonable and expected, because in the slowly rotating
approximation, the effect of rotation is too weak to change the
star's shape (almost spherical), which has strong influence on the
moment of inertia.

\section{Conclusion and Discussion }
As shown by Gupta et. al.\cite{s22}, the parameters of the EOS in
the QMDTD model have few influences on the configuration of the
strange star. We extend their work to the rotational case and have
studied the slowly rotating strange star with uniform angular
velocity, employing the QMDTD model and the Hartle's method. The
M-R relation , the moment of inertia and the frame dragging of
strange star with different rotating frequencies are given. We
found that mass, the moment of inertia and the frame dragging
increase when temperature increases. We also found that the
challenges of Stella and Vietri for HBO and the moment of inertia
$I_{45}/(M/M_ s>2.3)$ for five bright sources being neutron stars
also exist for the strange star. We cannot use the strange star to
solve this difficulty.

Furthermore, we have extended the Hartle's method to study the
differential rotating strange star with $\Omega=\Omega(r)$. We
found that the massive strange star can be prepared in the
differential rotation. In the slowly rotating approximation,
compared to the temperature the differential rotation has less
effect on changing the moment of inertia of the star.

\begin{acknowledgments}
This work was supported in part by NNSF of China under
NO.10375013, 10247001, 10235030, by the National Basic Research
Program 2003CB716300 of China and the Foundation of Education
Ministry of China under constact 2003246005.
\end{acknowledgments}


\end{document}